# Neural Network Based Next-Song Recommendation

Kai-Chun Hsu[1]     Szu-Yu Chou[2]     Yi-Hsuan Yang[2]     Tai-Shih Chi[1]

[1]Department of Electrical and Computer Engineering, National Chiao Tung University, Taiwan

[2]Research Center for IT innovation, Academia Sinica, Taipei, Taiwan

kch610596@gmail.com, {fearofchou, yang}@citi.sinica.edu.tw, tschi@mail.nctu.edu.tw

## ABSTRACT

Recently, the next-item/basket recommendation system, which considers the sequential relation between bought items, has drawn attention of researchers. The utilization of sequential patterns has boosted performance on several kinds of recommendation tasks. Inspired by natural language processing (NLP) techniques, we propose a novel neural network (NN) based next-song recommender, CNN-rec, in this paper. Then, we compare the proposed system with several NN based and classic recommendation systems on the next-song recommendation task. Verification results indicate the proposed system outperforms classic systems and has comparable performance with the state-of-the-art system.

## Categories and Subject Descriptors

• Information systems~Recommender systems

• Computing methodologies~Neural networks

## 1. INTRODUCTION

There are several types of recommendation tasks, and the task of next-item/basket recommendation emerged in recent years. This task is to predict the next item or the next basket of items which a user is interested in buying. Hence, successfully capturing the sequential buying pattern of the user is critical to this type of recommendation systems.

Existing algorithms for this task are mainly based on matrix/tensor factorization, Markov Chain (MC) or Markov Embedding [1-7]. It has been shown that combining the sequential pattern of buying behavior and the general taste of the user is more helpful to system performance than solely utilizing the sequential pattern or the general preference of the user [1, 2]. One might think this kind of recommendation system is equivalent to a hybrid of a conventional content-based (CB) recommendation system and a collaborative filtering (CF) recommendation system. However, a CB system only captures the similarity between items but not the sequential buying pattern of a specific user. Thus, for a specific user, we postulate that next-item/basket recommenders would be more welcomed than CB-CF hybrid systems. Conducting simulations to validate this assumption is beyond the scope of this study since we cannot obtain proper audio datasets for performance comparison.

A fundamental question to our assumption would be like "Does a user listen to music with some sequential pattern?". Empirically, people usually listen to songs of the same singer, album, music genre, lyricist, composer or record company in a listening session [4]. That is, people listen to related songs in a session. This phenomenon has been verified by embedding the listened songs of a user in a session into the Euclidean space [4, 5]. The study showed the positions of these songs are close to each other in the space while the positions of the songs listened in another session are far apart. The study also revealed that there is some correlation between songs in a session. In other words, the sequential pattern of listening behavior of a user does exist.

To the best of our knowledge, we are the first group to implement neural network (NN) based next-song recommenders. Firstly, we modified the next-basket recommender NN-rec developed in [8] to the next-song recommender. The original idea of the NN-rec was inspired by natural language processing (NLP) techniques [9]. The authors of NN-rec modified the neural probabilistic language model, which is used to predict the next word given all the previous words, to build the next-basket recommender by replacing the local context of a sentence with baskets of items and the user identity. The NN-rec has shown superior performance on two retail datasets [8], yet, it has not validated on music recommendation task. Secondly, also inspired by NLP techniques [10], we proposed a novel convolutional neural network (CNN) based recommender, the CNN-rec. In [10], the original deep learning architecture was modified by including a convolutional layer to capture the relation between adjacent words in sentences for re-ranking short text pairs. This technique has been corroborated in the field of NLP [10-14]. In the proposed CNN-rec, we adopted a convolutional layer to see whether the local relation between sequential songs can further improve the performance of the recommender. Thirdly, we implemented the Word2Vec method [15], which is also based on NN, as one of the baseline methods. One of the core techniques of these three NN-based methods is word embedding, which has been widely used in the field of NLP [9-14, 16] and successfully adopted in recommendation systems recently [8, 17]. In this paper, we investigate the efficacy of these three NN-based next-song recommenders and compare them with other two classic methods. Moreover, we investigate the impact of the number of songs kept in the historical listening sequence to system performance. To reproduce results of this paper, source codes of the proposed CNN-rec and other compared recommenders, whose codes are not publicly available, can be found on our website[1].

The rest of this paper is organized as follows. A brief introduction of our proposed system and compared systems are presented in Section 2. Section 3 demonstrates the experiment results. Lastly, we draw the conclusion in Section 4.

---

[1] Source codes of the used systems in this paper are available at https://github.com/huHHhhuhu/CNN-rec

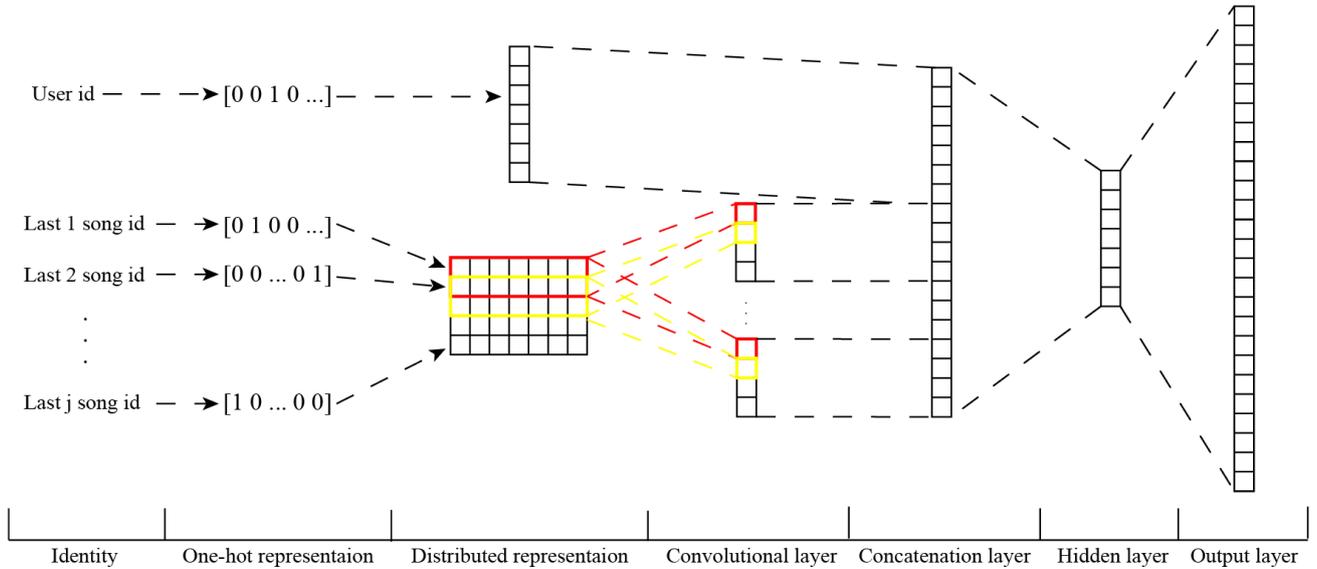

**Figure 1. Schematic diagram of the proposed CNN-rec**

## 2. PROPOSED AND COMPARED METHODS

### 2.1 Formalization of next-song recommendation

Most of next-song recommenders predict the next song for a user based on the last couple songs or one song listened by him. Each user has a listening record, which can be expressed as $I_u = (i_u^1, i_u^2, ..., i_u^t)$, where $i_u^t$ is the index of the song listened by the user $u$ at time $t$.

### 2.2 Proposed CNN-rec

Figure 1 shows the computational structure of the proposed CNN-rec recommender. Inspired by [10-14], we adopted a convolutional layer to capture the local relation pattern between adjacent songs.

The CNN-rec can be adjusted to consider any number of last songs listened by the user. Here, we use the last j songs as an example. First of all, we converted the identity of the user and the last j songs in his listening record to one-hot representations, which are high-dimensional highly sparse vectors, by 1-of-N encoding. Each user or song has a specific non-zero entity in the feature vector. Such kind of representation is widely used in the field of NLP [9-14, 16].

Secondly, we used the word embedding technique, commonly used in NLP research [9-16], to gain a low-dimensional dense feature. This embedded feature is referred to the distributed representation. The embedding process uses a learned mapping matrix to embed one-hot representations into the Euclidean space using the inner product of the one-hot representations and the mapping matrix. The embedded feature has the merit of utilizing the similarity between each other to reduce data sparseness for later processing. It is because once a system has been trained with an embedded feature, it is equivalently trained with similar embedded features. The embedding process also reduces the time/space complexity for later processing by producing a lower dimensional representation than the one-hot representation. Detailed explanations of the embedding process can be accessed in [9].

Thirdly, we stacked the song index features sequentially to form a matrix, which was put through the convolutional layer.

In a typical CNN, the convolutional layer is followed by a pooling layer. However, we didn't implement the pooling layer in our system because it reduced the performance in our experiments. Theoretically, we don't need to downsample the output of the convolutional layer because the fuzzification about the feature position is unnecessary in our case. Besides, the information of feature may be reduced in the downsampling process. Consequently, we directly concatenated all the song features from the output of the convolutional layer with the embedded user index feature to form a high-dimensional feature, as shown by the concatenation layer in Figure 1. Next, a hidden layer with a nonlinear activation function was placed after the concatenation layer.

Finally, the output layer with the softmax activation function produces the probability of each song being the next listened one.

### 2.3 Compared methods

Here we describe a set of compared methods.

• **NN-rec:** Inspired by the NN-based probabilistic language model in the field of NLP [9], the NN-rec was originally proposed as a next-basket recommender [8] with remarkable performance on two retail datasets [9].

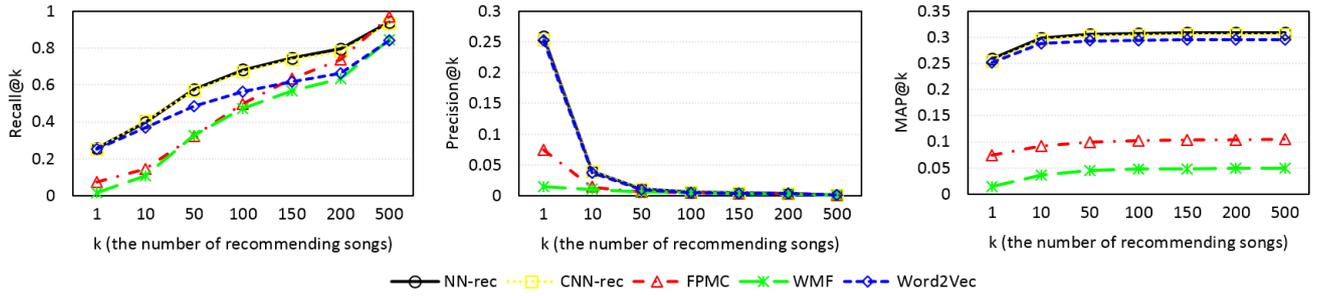

Figure 2. Performance of five compared recommenders

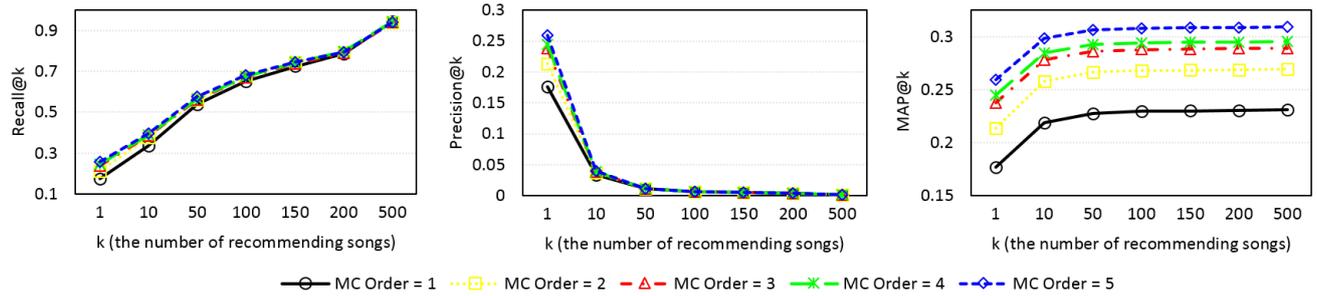

Figure 3. Impact of order of Markov chain to NN-rec performance

- **WMF:** The weighted matrix factorization (WMF) [18] is a state-of-the-art matrix factorization method which uses implicit feedback datasets. It estimates the confidence of preference by counting the number of request to the same song. In this paper, this method was treated as a representative of classic recommenders, which only consider users' general preference but not their sequential patterns.

- **Word2Vec:** The Word2vec method utilizes either the continuous bag-of-words (CBOW) or the continuous skip-gram architecture [15]. Both architectures contain a shallow neural network to perform word embedding and reconstructing linguistic contexts of words. Therefore, it can be used on recommendation [17]. In this paper, this method was selected as a representative of sequential recommenders, which only consider users' sequential patterns but not their general preference.

- **FPMC:** The factorizing personalized Markov chains (FPMC) was proposed to combine the common Markov chain with the matrix factorization technique [1]. The method produces a 3-dimensional tensor in the space of items. In this paper, we implemented this state-of-the-art method as a representative of combined recommenders, which consider both users' general preference and their sequential patterns.

## 3. EXPERIMENTS
### 3.1 Data Pre-processing
In our experiments, we used the last.FM-1k music dataset [19], which has been widely used in music recommendation experiments. It is the only open music dataset with timestamp records, which can be used to evaluate next-song recommenders. Its timestamps span from Feb. 2005 to May 2009. The dataset comprises 992 users, 176,948 artists, 1,505,264 items and 19,150,868 listening records.

As for data pre-processing, we first extracted top-10000 songs from the dataset to reduce noise, data sparseness and enormous computational requirements. Then we shuffled the data and partitioned the data into a training, a validation and a test set using the ratio of 7:1:2. Further, to generalize the prediction results, we deleted those records, which contain songs listened by users in the training set, from the validation and the test set. After the pre-process, we obtained a subset of data which comprises 987 users, 10000 songs and 4,086,781 records.

### 3.2 Validation Stage
In the validation stage, we fine-tuned NN-rec and CNN-rec systems to achieve their best performance for fair comparisons.

As for the structure of the NNs, the dimension of the embedding feature was set to 60 and the last 5 songs were considered for predicting the next song. There were 300 neurons in the hidden layer with ReLU activation function. In addition, the number of convolutional filters of CNN-rec was set to 325; the size of each filter was 2; and the filter stride was set to 1. The neurons in the convolutional layer were also with ReLU activation function.

During training, we used cross-entropy as the cost function and Adagrad [20] with backpropagation as the update rule for each method. The number of epoch was set to 25; the batchsize was set to 50 and the learning rate was 0.01. We used Glorot weight [21] for initialization and the dropout ratio of 0.7 in these two NN-based systems. It respectively takes about 80 minutes to train the NN-rec and 180 minutes to train the CNN-rec.

Other methods were also fine-tuned for comparisons. We found the Word2Vec achieved higher performance using skip-gram than CBOW, so the results of Word2Vec shown in this paper were obtained using the skip-gram architecture.

## 3.3 Evaluation

### 3.3.1 Performance Comparison

For evaluations, we used the evaluation metrics proposed in [22], which are more suitable for evaluating top-N recommendation task.

According to [4], the sequential relation between songs breaks once the continual listening pauses for a while. Hence, the recommenders should only learn sequential patterns within a certain period of time. Therefore, we divided the sequential listening records into sessions in which the inter-event time is less than an hour. In our opinion, the 1-hour interval we chose fits the music listening behavior of regular users.

Performance of all compared systems is shown in Figure 2. Clearly, we can see that all the NN-based next-song recommenders, i.e., NN-rec, CNN-rec and Word2Vec, outperform all the non-NN based systems except the recall rate at k=150, 200 and 500, where FPMC achieves higher performance (recall rate = 0.635, 0.7378 and 0.9686) than Word2Vec (recall rate = 0.6182, 0.6628 and 0.843). However, results at lower k (the number of recommending songs) are more important so that we still can say NN-based recommenders perform better than FPMC. Hence, we can conclude that neural networks are suitable for this task. Among NN-based recommenders, the NN-rec performs the best in all evaluations, the CNN-rec is always the second best and the Word2Vec is the worst one. These results show the NN-based recommenders, which consider users' general preference as well as their sequential listening patterns, can outperform the systems, which only considers the sequential listening patterns of users. Not meeting our expectation, the proposed CNN-rec is slightly worse than the NN-rec in terms of accuracy. One possible reason is that the local relations between adjacent songs provide no more information than the sequential patterns of the last 5 songs such that the CNN-rec shows no additional performance boost.

### 3.3.2 Impact of the Order of Markov chain

Performance of next-item/basket recommenders may depend on the order of Markov chain. For instance, FPMC [1], PME [5], PRME [7] and Bi-gram [24] are first-order Markov models while HRM [2], Tribeflow [6], NN-rec [9] and our CNN-rec can predict the next item using the last j items. But what is the impact of the order number in music recommendation task? To answer the question, we conducted experiments using the same NN-rec with different order numbers and the results are shown in Figure 3.

The results show that increasing the order number can improve the performance especially in situations where only a few items are recommended to users (i.e., when k is small). Nevertheless, we can see that the performance will gradually saturate as the order number increases. People probably listen to numerous similar songs continually, so considering the sequential patterns in a longer period is helpful to performance if we have enough data without overfitting.

## 4. CONCLUSION

We are the first group to implement and evaluate NN-based next-song recommendation systems. We also propose the CNN-rec, whose structure is transplanted from NLP, to catch local relations between adjacent songs using a convolutional layer. In addition, we show that the NN-based next-song recommenders, CNN-rec, NN-rec and Word2Vec, outperform the non-NN based ones. Our results demonstrate that the NN-based next-song recommenders, which combine users' general preference and sequential listening patterns, have the highest performance. Although the proposed CNN-rec does not outperform the NN-rec, which is a shallow NN, we think it can replace a deep-NN based recommender in situations where the time and space resources are limited. At the end, we show that increasing the order of Markov chain can gradually increase the performance till it plateaus.

## 5. ACKNOWLEDGMENTS

This research is supported by the Ministry of Science and Technology, Taiwan under Grant No MOST 103-2220-E-009-003.